\documentclass[prl,aps,amsmath,amssymb,groupedaddress,twocolumn]{revtex4}

\usepackage{graphicx}

\newcommand{\nc}{\newcommand}
\nc{\beq}{\begin{equation}}
\nc{\eeq}{\end{equation}}
\nc{\beqa}{\begin{eqnarray}}
\nc{\eeqa}{\end{eqnarray}}

\def\gsim{\mathrel{\rlap{\lower4pt\hbox{\hskip1pt$\sim$}}
    \raise1pt\hbox{$>$}}}       

\begin{document}

\title{Macroscopic superpositions and black hole unitarity}

\author{Stephen~D.~H.~Hsu} \email{hsu@msu.edu}
\affiliation{Department of Physics and Astronomy \\ Michigan State University  }

\begin{abstract}
We discuss the black hole information problem, including the recent claim that unitarity requires a horizon firewall, emphasizing the role of decoherence and macroscopic superpositions. We consider the formation and evaporation of a large black hole as a quantum amplitude, and note that during intermediate stages (e.g., after the Page time), the amplitude is a superposition of macroscopically distinct (and decohered) spacetimes, with the black hole itself in different positions on different branches. Small but semiclassical observers (who are themselves part of the quantum amplitude) that fall into the hole on one branch {\it will miss it entirely} on other branches and instead reach future infinity. This observation can reconcile the subjective experience of an infalling observer with unitarity. We also discuss implications for the nice slice formulation of the information problem, and to complementarity.
\end{abstract}


\maketitle

\date{today}

\bigskip


A recent paper by Almheiri, Marolf, Polchinski and Sully (AMPS) \cite{AMPS} (see also \cite{Mathur}) has stimulated new interest in the black hole information problem. AMPS study the entanglements of radiation, near-horizon modes and an infalling observer, concluding that unitarity of the evaporation process requires that the spacetime near the horizon deviates strongly from the vacuum state (exhibits a firewall), in violation of the equivalence principle. In this note we emphasize the importance to this discussion of decoherence and superpositions of macroscopically different states. We believe such considerations resolve the paradox proposed by AMPS and allow for unitarity without requiring firewalls. Further, these considerations may reveal the mechanism for unitarity itself. For related discussions, see \cite{HsuBH, Nomura, Raju, Hayden}.

Consider quantum excitations, including gravitons, on a flat spacetime background. Assume that the initial quantum state $\Psi_i$ at past infinity is a semiclassical configuration of matter or radiation (e.g., a dustball, or some energetic photons) which will evolve into a large black hole $B$ through gravitational collapse. $\Psi$ may also describe some additional degrees of freedom (``observers'') whose trajectories carry them into the vicinity of $B$ after it has formed. We assume $B$ undergoes complete Hawking evaporation, so the final state $\Psi_f$ at future infinity describes outgoing radiation plus any observers that do not cross the horizon into the hole. The information problem in this scenario is equivalent to whether $\Psi_i$ and $\Psi_f$ are related by a unitary transformation. The puzzle is motivated by the fact that the Hawking radiation originates in a region which is (at least semiclassically) causally disconnected from objects which are behind the horizon. How does the quantum information associated with these objects reach future infinity?

Let us examine the time evolution of $\Psi$ in detail. It is well known \cite{Page} that the evaporation of $B$ in the usual semiclassical approximation eventually leads to significant uncertainty in both its position and momentum. When a fraction $f$ of the total rest mass has evaporated, the uncertainty in position is $\Delta x \sim f^{3/2} M^2$. At the Page time $t_{\rm Page} \sim M^3$, $\Delta x \sim M^2$, which is much larger than the size of $B$. This is not surprising, as even partial evaporation of a black hole converts a nontrivial fraction of its rest mass into a large number of uncorrelated radiation quanta, each producing recoil impulses on $B$ due to conservation of momentum.

All of the possible sequences of radiation impulses are described by (contained in) the Schrodinger evolution of $\Psi$. If two different radiation sequences cause the position of $B$ to differ macroscopically, those two branches of the wave function decohere from each other: the position of the black hole is entangled with (``measured by'') environmental degrees of freedom, such as gravitons. Two decohered branches will subsequently have very small influence on each other, but nevertheless remain part of the global wave function $\Psi$ \cite{MW}.

Thus, at intermediate times (e.g., $t > t_{\rm Page}$), $\Psi$ must be a superposition of decohered branches, each describing a macroscopically different situation: $B$ in different locations, and with different velocities (see figure). Now consider a semiclassical observer Alice, small in size relative to $B$, with a well-defined trajectory that passes close to $B$. As we stressed above, Alice is present in the initial state $\Psi_i$, and all of her possible outcomes (including those resulting from ``choices'' she makes about, e.g., how to fire her rockets \cite{Alice}) are described by the unitary evolution of $\Psi$ (see \cite{HsuQM} for a discussion of determinism and free will in unitary quantum mechanics). Does Alice fall into the black hole? Even if on some branches her (semiclassical) trajectory intersects $B$, there are many branches on which it does not. Alice is by assumption much smaller than $B$, and the uncertainty in the position of $B$ after the Page time is of order its Schwarzschild radius squared: $\Delta x  \sim R^2$. Even if Alice's trajectory intersects $B$ on one branch, it does not intersect most of the $\Delta x^3 \sim R^6$ volume where $B$ is likely to be found on other branches.

\begin{figure}[h]
\includegraphics[width=8cm]{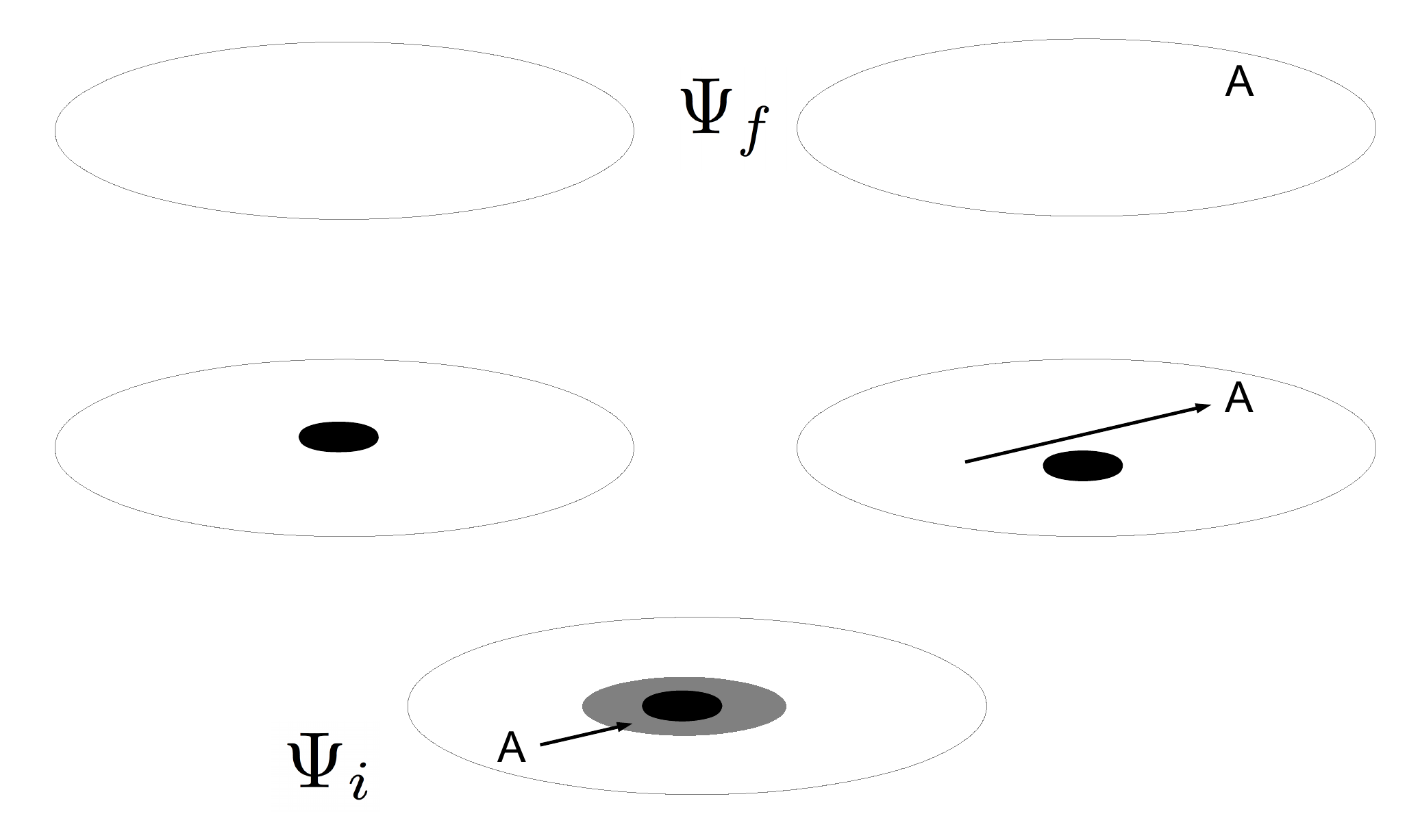}
\caption{Schrodinger evolution of initial state $\Psi_i$, which describes a black hole progenitor and Alice. At intermediate times ($t > t_{\rm Page}$) $\Psi$ has ({\it unavoidably} -- due to radiation recoil) decohered into two branches with the black hole in different locations. On one of these branches Alice falls into the hole, which subsequently evaporates. On the other branch Alice does not fall into the hole and consequently is still present in $\Psi_f$.}
\label{figure}
\end{figure}

Therefore, the final state $\Psi_f$ contains branches in which Alice reaches future infinity, as well as ones on which Alice has fallen into the black hole and her mass has been converted into Hawking radiation. As we discuss below, this is also plausibly true for observers that fall into the hole before the Page time (i.e., before the uncertainty in the location of $B$ is macroscopic), and even for the progenitor particles that formed $B$ at early times. A macroscopic uncertainty in the black hole position and velocity, corresponding to decoherent branches of $\Psi$, makes it rather obvious that $\Psi_f$ contains versions of Alice which never fell into $B$. But the conclusion applies as well to {\it any} information-carrying degrees of freedom which have some probability (even exponentially small) of escaping or not falling into $B$.

Now consider the firewall question. A particular semiclassical instantiation of Alice -- one with memory records of a particular location for $B$ and its associated spacetime -- cannot detect the other decohered Alices without resources which are exponential in the black hole entropy \cite{HsuBH}. Consequently, Alice cannot measure the global state of the Hawking radiation: it is spread over all of the decohered branches (this is also emphasized by Harlow and Hayden \cite{Hayden} from the perspective of quantum computation). Thus it is plausible that the AMPS paradox is not realized. To proceed further, adopt momentarily the following notation: $B$ = near-horizon modes of an old black hole, $R$ = early Hawking radiation and $A$ = interior modes of the black hole. The AMPS observation is that (1) the equivalence principle, applied to Alice as she falls through the horizon, requires nearly maximal $BA$ entanglement (Alice sees the Minkowski vacuum at the horizon) and (2) nearly maximal $BR$ entanglement is required by unitarity. These two requirements are contradictory, leading to a firewall (deviation from vacuum state at the horizon) if we keep (2) and drop (1). However, the problem with the argument is that the $B$ of (1) and (2) are different. The $B$ of (1) which is observed by {\it a particular semiclassical instantiation of Alice} refers specifically to the near horizon modes {\it on her branch}. The $B$ and $R$ of (2) are spread across many macroscopically different spacetimes (branches), so the basic formulation of the firewall paradox is problematic.

What about a particle that enters the black hole $B$ at early times (e.g., before the Page time) or a progenitor particle from which the black hole is formed? Are there also branches of $\Psi_f$ on which this particle reaches future infinity? Page \cite{Page} suggested two plausible mechanisms which could make this possible. The first involves ordinary quantum mechanics: to localize the wave function of a particle completely within a finite spatial region requires infinite momentum components (this is a simple consequence of Fourier analysis). Thus it is plausible that even the individual progenitor particles have small amplitudes to be found outside the black hole; heuristically, they tunnel out the hole as it is formed. A second possibility is that small violations of semiclassical causality, or locality, may be intrinsic to quantum gravity. For example, there may be amplitudes of size $\exp (- R^2) \sim \exp (- S_B)$ for a particle to escape a black hole of radius $R$. These mechanisms could lead to branches of $\Psi_f$, albeit of exponentially small amplitude, containing {\it in aggregate} all of the quantum information which, on the largest amplitude branches of the wave function, fell into $B$. The existence of such branches was discussed in \cite{HsuBH} and the possibility that they might unitarize (purify) the Hawking radiation state was discussed in \cite{Raju}. Although the amplitudes in question are small, the correspondingly large dimensionality ($\exp S_B$) of the radiation Hilbert space provides a compensating factor. Note that a pure radiation state measured by an observer with limited experimental precision would appear to be a thermal mixed state \cite{HsuBH}.

Plausibly, there is enough information in $\Psi_f$ to reconstruct the initial quantum state: the evolution $\Psi_i \rightarrow \Psi_f$ could be unitary, even if a black hole is formed at intermediate times. Heuristically, the time-reversed Schrodinger evolution of $\Psi_f$ results in $\Psi_i$, because many ($\sim \exp S_B$) branches of $\Psi_f$, each containing partial aspects of the quantum information that fell into $B$, interfere perfectly to produce the original initial state.

\bigskip

Perhaps the clearest formulation of the information problem utilizes the so-called nice slice (a spacelike slice intersecting both infalling matter and outgoing radiation) and the quantum no cloning theorem \cite{Mathur}. One concludes that the quantum information associated with an object behind the horizon cannot also be found in the outgoing radiation. However, by formulating the problem in terms of the Schrodinger evolution of $\Psi$ as we have, the gap in the nice slice argument is revealed: a nice slice is defined on a particular semiclassical spacetime, but the evolution of $\Psi$ {\it inevitably} generates a superposition of different spacetimes, with the black hole in different locations. The Penrose diagram on which the nice slice is drawn {\it cannot also describe} the other branches of $\Psi$ on which the object depicted as behind the horizon (the one whose quantum state cannot be cloned in the outgoing radiation) has {\it instead avoided entering the hole} and is itself part of the outgoing matter.

An old question: How do we reconcile Alice's subjective experience (in particular, preserving the equivalence principle or ``no drama'') with unitarity? The answer is that the branch on which Alice experiences falling through the horizon does not, by itself, constitute a unitary evolution of $\Psi_i$. Rather, it is $\Psi_f$, the superposition of all branches, including ones in which Alice does {\it not} fall into the hole, that represents the unitary evolution of $\Psi_i$. There is no contradiction between a particular subjective experience (e.g., on a specific spacetime to which the nice slice argument applies, obeying the equivalence principle) and overall unitarity.

Where is complementarity? It can be elucidated by comparing the perspective of a particular Alice who falls into the hole with that of a super-observer whose observations can overcome decoherence. (This generally requires either exponential experimental sensitivity or the ability to make non-local measurements ``all at once'' \cite{HsuBH,Hayden}.) There is no requirement that the subjective experience of {\it a particular} Alice be consistent with the measurements of a super-observer who is aware of {\it all of the Alices} and, hence, can verify unitarity. Despite the central role of gravity in black hole physics, the complementarity that emerges here is related to the original one proposed by Bohr in the context of quantum measurement: the wave or particle nature of a photon is determined by the specific measurement process used to probe it. The apparent fate of black hole information depends on whether the observer can detect the entire global state or merely a particular branch of $\Psi$. The super-observer cannot fall into the black hole, and Alice cannot determine the global state $\Psi$. 

In ordinary quantum mechanics, a super-observer can verify that wave functions evolve unitarily (rather than collapse), whereas to an observer of limited measuring capability pure states {\it appear} to evolve into mixtures \cite{Bell}. Similarly, a super-observer of black hole formation and evaporation could verify that $\Psi$ evolves unitarily. To an observer of limited measuring capability (e.g., Alice) information {\it seems} to be lost behind the horizon, and black holes {\it appear} to cause pure states to evolve into mixtures.


\bigskip

\emph{Acknowledgements---}  The author thanks Dieter Zeh for prior correspondence concerning black holes, decoherence and unitarity. This work was supported in part by the Office of the Vice-President for Research and Graduate Studies at MSU, and by the Department of Energy under DE-FG02-96ER40969.


\bigskip



\begin{thebibliography}{99}
\bibitem{MW} Some readers may be uncomfortable with the many worlds flavor of our discussion. We consider unitary evolution of $\Psi$ without collapse (von Neumann projection), and the subjective experiences of (many versions of) observers described by $\Psi$. In the traditional Copenhagen interpretation, measurement consists precisely of a pure state evolving into a mixed state -- each element of the mixture describing one of the mutually exclusive outcomes. Hawking, in his original formulation of the black hole information puzzle, emphasized that in the semiclassical approximation black hole evaporation causes a pure state to evolve into a mixed state, which he deemed problematic. An interpretation that does not allow pure states to evolve into mixed states (other than FAPP -- see \cite{Bell} below) is a many worlds interpretation.
\bibitem{Alice} If Alice steers her rocket adaptively to follow the black hole (regardless of where radiation recoils take it), she can arrange to enter it on the branches of largest amplitude, assuming her steering and information processing are reliable. Note though that the memory records of the different Alices entering black holes at different positions will not coincide -- they will recall different steering adjustments and spatial coordinates. In addition, there will still be amplitudes of small size -- presumably no smaller than the $\exp( - S_B)$ amplitudes discussed in the main text -- on which she avoids falling into the hole.
\end{thebibliography}
\end{document}